\documentclass[10pt,
               aps,
               pre,
               twocolumn,
               showpacs, 
               groupedaddress]{revtex4-2}
\usepackage[utf8]{inputenc}
\usepackage{graphicx}
\usepackage[caption=false]{subfig}
\usepackage{amsmath, amsfonts, amssymb}
\usepackage{hyperref}
\usepackage{tikz}
\makeatletter
\newcommand*{\rom}[1]{\expandafter\romannumeral #1}
\makeatother

\begin{document}

\title{Viscoelastic response of an active particle under the action of magnetic field}

\author{M Muhsin}

\affiliation{Department of Physics, University of Kerala, Kariavattom, Thiruvananthapuram-$695581$, India}

\author{F Adersh}

\affiliation{Department of Physics, University of Kerala, Kariavattom, Thiruvananthapuram-$695581$, India}

\author{M Sahoo}
\email{jolly.iopb@gmail.com}
\affiliation{Department of Physics, University of Kerala, Kariavattom, Thiruvananthapuram-$695581$, India}

\date{\today}

\begin{abstract}
 We consider the dynamics of a charged inertial active Ornstein-Uhlenbeck particle in a viscoelastic suspension under the action of an uniform magnetic field. With the help of both numerical simulation and analytical framework, we exactly investigate the viscoelastic response of the particle to the magnetic field by means of particle trajectories, mean displacement, and mean square displacement (MSD) calculations. The simulated particle trajectories and MSD calculations reveal that the steady state response of a confined harmonic particle to the magnetic field show interesting features due to the complex interplay of underlying physical processes such as elastic dissipation and active fluctuations in the medium. When the activity time scale of the dynamics is larger than the elastic dissipation timescale (or vice-versa), the steady state MSD shows an enhancement (or suppression) with increase in the field strength. In both the cases, for very strong magnetic field the MSD interestingly approaches the value at the equilibrium limit where the generalized fluctuation dissipation relation is satisfied and the the particle behaves like an inertial passive Brownian particle. Thus, for a finite magnetic field, the system exhibits a re-entrant type transition from active to passive and then to active behaviour with persistence of elastic dissipation in the medium. However, for a free particle, the overall displacement covered always gets suppressed with increase in the strength of the magnetic field and finally approaches zero for a very strong magnetic field.
\end{abstract}  

\maketitle
\section{INTRODUCTION}\label{sec:intro}
In the recent years, the study of active matter has received considerable attention in different branches of science~\cite{ bechinger2016active, Ramaswamy2017active,Gompper2020roadmap, pietzonka2021oddity, magistris2015intro}. The term ``active matter" refers to a class of soft-matter systems where its individual constituents are capable of making autonomous movement by consuming energy from the environment. The motion governed by these constituent particles is called as active Brownian motion or self-propelled motion. Active matter systems are immensely pursued both theoretically as well as experimentally~\cite{dauchot2019dynamics, bashirzadeh2019encapsulation, scholz2018rotating, andac2019active, makarchuk2019enhanced, taktikos2014how, aranson2022bacterial, needleman2017active, vernerey2019biological,bechinger2016active}. In particular, the biological active matter is significant because of its complex dynamics in crowded and stochastic environments~\cite{taktikos2014how,aranson2022bacterial}.
Similarly, the artificially synthesized active matter systems like collection of Janus particles~\cite{walther2013janus, howse2007self}, living crystals~\cite{palacci2013living}, and swarming robots~\cite{wang2021emergent} have wide applications in the areas of
robotics~\cite{aguilar2016review}, material science,
and nano-medicines~\cite{arijit2020active}.

From the theoretical perspective, a variety of models have been used to describe the dynamics of active matter. Among them, the active Brownian particle (ABP) model~\cite{hagen2009NonGaussian, hagen2011brownian, cates2013when, kanaya2020steady, buttinoni2013dynamical, fily2012athermal, stenhammar2014phase, bialke2015negative, solon2015pressure, caprini2021collective, caprini2020hidden, roon2022role, scholz2018inertial, mandal2019motility} and active Ornstein-Uhlenbeck particle (AOUP) model~ \cite{lehle2018analyzing, bonilla2019active, martin2021statistical, caprini2019active, caprini2021inertial, dabelow2019irreversibility, berthier2019glassy, wittman2018effective, fily2019self, mandal2017entropy, fodor2016far} are the two most successful ones in explaining many interesting aspects of the self-propelling systems~\cite{marini2015towards, Gompper2020roadmap, cates2015motility}. In most of these studies, the particles are considered to be suspended in a Newtonian bath with an instantaneous friction and with no memory effect of the medium or environment. However, in many situations, the particles are exposed to non-Newtonian environments like polymer solutions~\cite{martinez2014flagellated}, crystalline domain~\cite{zhou2014living}, and biological surroundings such as mucus~\cite{sleigh1988propulsion}, tissues etc with finite memory effect of the medium or environment. The natural active systems are mostly non-Newtonian and are often found to be viscoelastic in nature with a transient elasticity in the medium. Such systems can be modeled using the generalized Langevin equation which includes the delayed memory effects of the medium in the viscous term. Therefore, it is crucial to study the viscoelastic effects on the dynamics of such systems as the memory effect of the medium cannot be ignored~\cite{patteson2015running, schamel2014nano, solano2016dynamics, narinder2018memory, sprenger2022active, sevilla2019generalized}. 

Active Brownian particle in a Newtonian bath, when subjected to an external magnetic field exhibits exotic transient and steady state properties~\cite{sandoval2016magnetic,fan2017ellipsoidal, vuijk2020lorentz,abdoli2021stochastic,muhsin2022inertial,muhsin2021orbital}. 
In our previous work, we observed that a harmonically confined active Ornstein-Uhlenbeck particle (AOUP), while self-propelling in a viscoelastic environment in the presence of a magnetic field features a novel phase transition from diamagnetic to paramagnetic phase~\cite{muhsin2021orbital}. In the present work, we employ this model to investigate the transport behaviour of an inertial active OU particle in a viscoelastic suspension characterized by an exponentially decaying memory kernel in the viscous drag. The particle is charged and subjected to a constant magnetic field perpendicular to the plane of its motion~\cite{Sahoo2007charged,kumar2012classical,kumar2012erratum, paraan2008brownian,lisy2013brownian}. We simulate the dynamics and obtain  instantaneous trajectories of both free and harmonic particle. From the exact solution of the dynamics, we derive the analytical results for the mean displacement (MD) and MSD. Here, we consider the corresponding time scales of activity or self propulsion and elastic dissipation not to be same so that the generalized fluctuation dissipation relation(GFDR) breaks down. Depending on the complex interplay of these two time scales, our analysis reveal that the steady state response of a harmonically confined particle to the magnetic field is interesting and it shows opposite behaviour in two different circumstances. In particular, when the persistence duration of activity in the medium dominates over elastic dissipation, the steady state MSD shows enhancement and when the persistence of elastic dissipation dominates over activity or self propulsion, it shows suppression with increase in strength of magnetic field. Moreover, for a very strong magnetic field, the MSD approaches the equilibrium value at which the GFDR holds good and becomes independent of both these time scales, reflecting the passive nature of the particle. 

In sec \ref{sec:model}, we introduce our model, define the physical parameters of interest and explain the simulation details. In sec.~\ref{sec:result}, we discuss our results and findings both from simulation and analytical calculation, which are followed by a summary in sec.~\ref{sec:summary}.

\section{MODEL}\label{sec:model}
\begin{figure*}[!ht]
    \centering
    \includegraphics[scale=0.682]{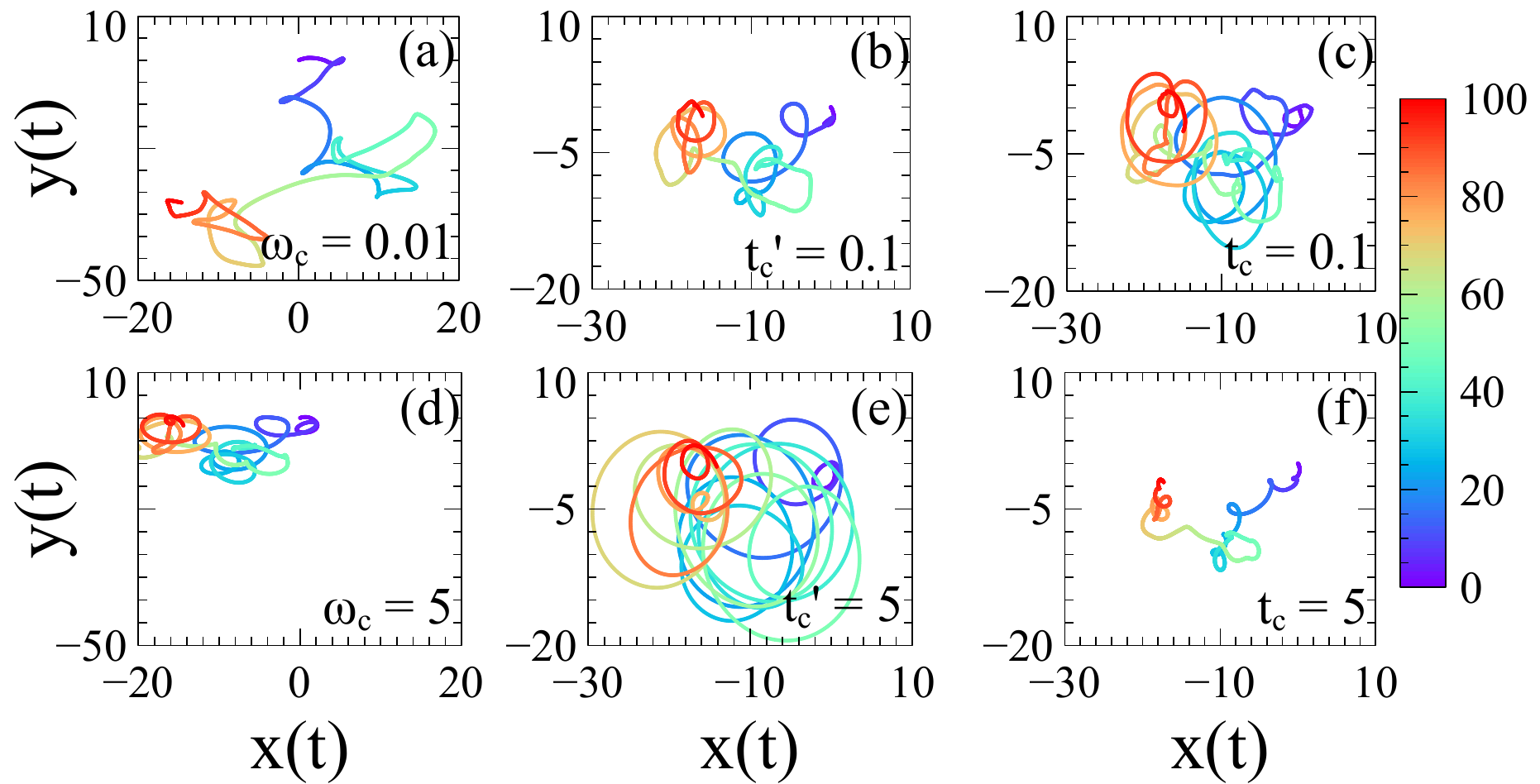}
    \caption{Simulated trajectory of a free particle for two different values of $\omega_c$ in (a) and (d), for two different values of $t_c^\prime$ in (b) and (e), and for two different values of $t_c$ in (c) and (f), respectively. The color map denotes the time evolution of trajectory. $t_c = t_c'$ is considered as unity in (a) and (d), $\omega_c = t_c$ is considered as unity in (b) and (e), and $\omega_c = t_c'$ is considered as unity in (c) and (f), respectively. The other common parameters are $\Gamma = 1$, $v_0 = 1$, $z_0 = 0$ and $D = 1$.}
    \label{fig:traj_free}
\end{figure*}
We consider the motion of an inertial active Ornstein-Uhlenbeck particle  suspended in a viscoelastic suspension. The particle is charged and constrained to move in a two-dimensional(2D) harmonic potential $U(x,y)=\frac{1}{2}k \left(x^2+y^2\right)$, with $k$ as the harmonic constant. Its motion is coupled to the presence of a constant magnetic field ${\bf B}=B\hat{z}$ applied perpendicular to the plane of motion. The dynamics is governed by the generalized Langevin's equation \cite{muhsin2021orbital, sevilla2019generalized} of motion

\begin{equation}
\begin{split}
  m\ddot{\bf X}(t) = & - \int_{0}^{t} f\left(t-t'\right) {\bf \dot{X}}(t')\, dt'+\frac{|q|B}{c}\sigma\cdot {\bf \dot{X}}(t) \\
  & - k {\bf X}(t) + \sqrt{D} {\mathbf{\xi}}(t).
    \label{eq:maineqmotion-vector}
\end{split}
\end{equation}

Here, ${\bf X} = \begin{pmatrix}
    x \\
    y
\end{pmatrix}$ refers to the position, $m$ stands for the mass and $\ddot{\bf X}(t)$ is the acceleration of the particle. The first term in right hand side of Eq.~\eqref{eq:maineqmotion-vector} is the drag force, that arises due to the interaction of the particle with the surrounding medium. Because of the elastic nature of the medium, the dynamics of the particle possess finite memory and is non-Markovian. Hence the drag force at particular instant of time not only depend on the velocity at that instant of time but also it depends on the velocities at previous times. Therefore, the drag force at a certain time $t$ is dependent on a friction kernel of the form
\begin{equation}
f\left(t-t'\right)=\frac{\gamma}{2 t_c'}e^{-\frac{t-t'}{t_c'}},\qquad t\geq t'
\end{equation}
with $\gamma$ being the friction or viscous coefficient of the medium. The friction kernel gives maximum contribution to the instantaneous velocity, and its contribution to the past velocities decays exponentially with a rate $\frac{1}{t_c'}$. $t_{c}^{'}$ is the time up to which the elastic dissipation sustains in the medium and for $t > t_{c}^{'}$, the viscous dissipation dominates. In $t_c' \to 0$ limit, the friction kernel becomes delta correlated and the system is left only with viscous dissipation\cite{muhsin2022inertial}. The friction kernel $f(t-t')$ follows the Maxwellian viscoelastic formalism as per which the elastic force of fluid  completely relax down to zero for sufficiently long interval of time and the suspension becomes viscous in nature \cite{muhsin2021orbital}. $\sigma [= \begin{pmatrix}
    0 & -1 \\
    1 & 0
\end{pmatrix}]$ is an arbitrary matrix and $\xi(t) [= \begin{pmatrix}
    \xi_x(t) \\
    \xi_y(t)
\end{pmatrix}]$ represents the Ornstein-Uhlenbeck noise followed by the moments
\begin{equation}
\langle \xi_\alpha(t) \rangle =0, \qquad  \langle \xi_\alpha(t)\xi_\beta(t') \rangle = \frac{\delta_{\alpha\beta}}{2t_c} e^{-\frac{|t-t'|}{t_c'}}.
\label{eq:noise_stat}
\end{equation}
Here, $(\alpha,\beta)\;\epsilon\;(X,Y)$ and $D$ is the strength of the noise. $t_c$ represents the noise correlation time or activity time of the dynamics. It is the time up to which the particle has a tendency to execute self-propulsion in the medium along a specific direction even with random collisions. The noise correlation in the dynamics decays exponentially with the time constant $t_c$. For a finite $t_{c}$ value, the system is in nonequilibrium even in the absence of memory kernel. However, the system approaches thermal equilibrium only when both the activity time scale and viscoelastic time scale are of equal order ($t_{c}=t_{c}^{'}$) and $D=2\gamma K_B T$. This is the condition at which the GFDR is validated. Because of the linearity of the model, one can solve Eq.~\eqref{eq:maineqmotion-vector} using Laplace transform method formalism. Introducing a complex number $z = x + iy$, Eq.~\eqref{eq:maineqmotion-vector} can be written as
\begin{equation}
   \ddot{z}(t)+\frac{1}{m}\int_{0}^{t} f(t-t') \dot{z}(t')\, dt' -i \omega_c \dot{z}(t) + \omega_0^2 z(t)=\epsilon(t),
   \label{eq:maineqmotion}
\end{equation}
where $\omega_c=\frac{|q|B}{mc}$, $\omega_0=\sqrt\frac{k}{m}$, and
$$\epsilon(t)=\frac{\sqrt{D}}{m}[\xi_x(t)+i \xi_y(t)].$$
The mean displacement (MD) can be obtained using the general definition as
\begin{equation}
    \langle R(t) \rangle= \langle z(t)-z(0) \rangle.
    \label{eq:md-def}
\end{equation}
Similarly, the MSD can be obtained as
\begin{equation}
    \langle R^2(t) \rangle= \langle |z(t)-z(0)|^2 \rangle.
    \label{eq:msd-def}
\end{equation}
Inorder to simulate the dynamics, Eq.~\eqref{eq:maineqmotion-vector} can be expressed in a Markovian form by introducing the variable ${\bf U}$ such that
\begin{equation}
    {\bf U}(t) = \int_{0}^{t} f\left(t-t'\right) {\bf V}(t')\, dt'.
    \label{eq:def_U}
\end{equation}
Then Eq.~\eqref{eq:maineqmotion-vector} can be written as 
\begin{align}
    \dot{X} &= V  \\
    m\dot{\bf V} &=  - {\bf U} +\frac{|q|B}{c}\sigma\cdot {\bf V} - k {\bf X} + \sqrt{D} {\mathbf{\xi}} \\ 
    \dot{{\bf U}} &= -\frac{{\bf U}}{t_c'} + \frac{\Gamma}{2 t_c'}{\bf V}
\end{align}
This set of Markovian equations can be simulated using the Heun's method formalism\cite{burrage2007numerical}. The Ornstein-Uhlenbeck noise ($\xi$) along with the properties given in Eq.~\eqref{eq:noise_stat} is implemented using Fox method approach as described in Ref.~\cite{fox1988fast}.
The simulation is run for $10^3$s with a time step of $10^{-3}$s. The steady state measurements are taken averaging over $10^5$ realizations and after ignoring the initial transients of $10^2$s for each realization. 
\section{RESULTS AND DISCUSSION}\label{sec:result}
\subsection{Free particle in the presence of magnetic field}\label{sec:free-mag}
For a free particle in the presence of magnetic field, the generalized Langevin equation of motion [Eq.~\eqref{eq:maineqmotion}] reduces to
\begin{equation}
   \ddot{z}(t)+\frac{1}{m}\int_{0}^{t} f(t-t') \dot{z}(t')\, dt' -i \omega_c \dot{z}(t) + =\epsilon(t).
   \label{eq:FPmg}
\end{equation}
By taking the initial conditions $z(0) = z_0$ and $\dot{z}(0) = v_0$, and using the Laplace transform method the solution of Eq.~\eqref{eq:FPmg} can be obtained as
\begin{equation}
  z(t)= z_0 + \sum_{i=0}^{2} a_i \int_{0}^{t} e^{s_i(t-t')}\xi(t') \,dt'+\sum_{i=0}^{2} a_i v_0 e^{s_i t},
\end{equation}
where $s_i$'s are given by \\
$s_0 = 0$ and
\begin{equation}
    \begin{split}
        s_{1,2}=-\frac{1}{2} \biggr[+\frac{1}{t_c'}-i  \omega _c \pm\frac{\sqrt{1-t_c' \left(2 \Gamma +\omega _c \left(\omega _c t_c'-2 i\right)\right)}}{t_c'}\biggr].
    \end{split}
\end{equation}
Here, $\Gamma = \frac{\gamma}{m}$ and the coefficients $a_i$'s are
\begin{equation}
    \begin{split}
        &a_0=\frac{1}{s_1 s_2 t_c'}\quad, a_1=\frac{s_1 t_c'+1}{s_1 \left(s_1-s_2\right) t_c'}\quad \text{and}\\
        &a_2=\frac{s_2 t_c'+1}{s_2\left(s_2-s_1\right) t_c'}\nonumber
    \end{split}
\end{equation}
In Fig.\ref{fig:traj_free}, We have shown the simulated particle trajectories for three different cases. Figs.~\ref{fig:traj_free} (a) and (d) represent the particle trajectories for a low and high strengthen magnetic field ($\omega_{c}$). It is noticed that under the influence of magnetic field the particle deviates from the motion along x-direction and gets dragged towards y-direction and has a tendency to make circular trajectories. When $\omega_{c}$ is large, these circular trajectories get squeezed and particle suffers a kind of trapping or suppression. As a consequence, the overall displacement of the particle reduces. This can be interpreted from Fig.~\ref{fig:traj_free} (d). In Figs.~\ref{fig:traj_free}(b) and (e), the particle trajectories are plotted for a small and large value $t_{c}^{'}$. It is observed that with increase in $t_{c}^{'}$, the particle trajectories get smoother and the particle makes more number of turns. At the same time on average it covers larger distances before completing the turns. Because of which, the circular trajectories get enhanced initially and subsequently reduces displaying a spiral kind movement. Similarly, the simulated particle trajectories for a small and large value of $t_{c}$ are shown in Figs.~\ref{fig:traj_free} (c) and (f), respectively. With increase in $t_{c}$ value, the particle makes more directed motion rather than making circular trajectories. At the same time, the particle becomes more confined, that is why the overall displacement decreases with longer persistent of activity in the medium.

Using Eq.~\eqref{eq:md-def}, MD for a free particle in the presence of magnetic field is calculated as
\begin{equation}
    \begin{split}
     \langle R(t) \rangle^{(f)} &= \langle z(t)-z(0) \rangle\\
       &= a_0v_0 + a_1v_0 e^{s_1t} + a_2v_0 e^{s_2t}.
    \end{split}
    \label{eq:md-free-expression}
\end{equation}
\begin{figure}[!ht]
\includegraphics[scale=0.757]{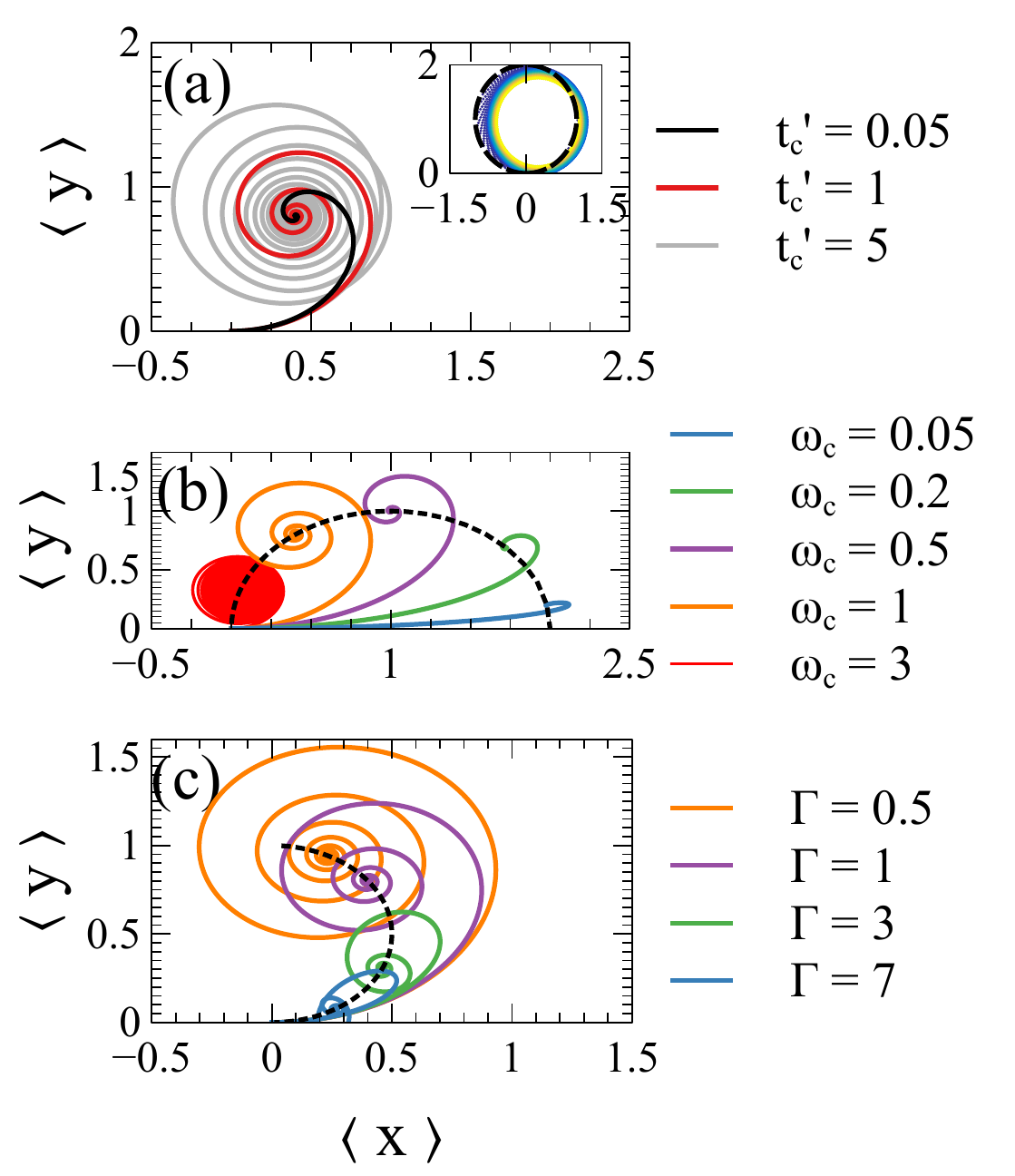}
\caption{The 2D parametric plots of MD [Eq.~\eqref{eq:md-free-expression}] is shown for different values of $t_c^\prime$ in (a), for different values of $\omega_c$ in (b), and for different values of $\Gamma$ in (c), respectively. The dashed lines in (b) and (c) represent the 2D evolution of MD with $\omega_{c}$ and $\Gamma$, respectively. Inset of (a) shows the evolution of MD for $t_c' = 20$ along with the dotted circle representing MD in the $t_c' \to \infty$ limit. Other common parameters are $D = 1$ and $\omega_c = 1$, $\Gamma = 1$ in (a), $t_c' = 1$, $\Gamma = 1$ in (b) and $t_c' = 1$, $\omega_c = 1$ in (c).}
\label{fig:md_free}
\end{figure}
Expanding Eq.~\eqref{eq:md-free-expression} around $t = 0$ in terms of powers of $t$, we obtain MD in the transient regime as
\begin{equation}
    \langle R(t)\rangle^{(f)} = v_0 t+\frac{i}{2}v_0\omega_c t^2 +O\left(t^3\right).
     \label{eq:md_trans_infty}
\end{equation}
In the time asymptotic limit ($t \rightarrow \infty$) or at steady state, $\langle R(t) \rangle$ reduces to
\begin{equation}
    \langle R \rangle^{(f)}_{st} = \lim_{t\to \infty}\langle R(t) \rangle^{(f)} =\frac{2v_0\left(\Gamma + i2\omega _c\right)}{4 \omega _c^2+\Gamma ^2}.
    \label{eq:md_free_infty}
\end{equation}

Hence, it is confirmed that the steady state MD has dependence neither on the elastic dissipation time nor on noise correlation time or activity time, rather it has dependence on the magnetic field as well as on viscosity of the medium or inertial time scale. The 2D time evolution of MD [$\langle y(t) \rangle$ vs $\langle x(t) \rangle$] for different values of $t_{c}'$, for different values of $\omega_{c}$, and for different values of $\Gamma$, are shown in Figs.~\ref{fig:md_free} (a), (b) and (c), respectively. For a given set of parameters, the MD evolves with time and saturates to a constant value in the long time limit, displaying a spira mirabilis kind trajectory \cite{sprenger2022active}. From Fig.~\ref{fig:md_free} (a), it is noticed that MD has a dependence on $t_{c}^{'}$ in the transient regime whereas the steady state MD is independent of $t_{c}^{'}$. Further, it is noteworthy that with increase in $t_{c}^{'}$ value, the particle on average takes longer time and covers larger distances before approaching the steady state at which the MD is saturated to a constant value. This fact concludes a kind of delay in the particle motion for approaching steady state because of the presence of memory in the viscous kernel and it represents a memory induced delay in the dynamics. Further, in $t_{c}' \rightarrow 0$ limit, the $\langle R(t)\rangle^{(f)}$ reduces to
\begin{equation}
	\langle R(t) \rangle^{(f)} = \frac{2 v_0}{\Gamma - 2 i\ \omega_c} \left( 1 - e^{-\frac{\Gamma t}{2} + i \omega_c\ t} \right),
\end{equation}
\begin{figure*}[!ht]
    \centering
    \includegraphics[scale=0.66]{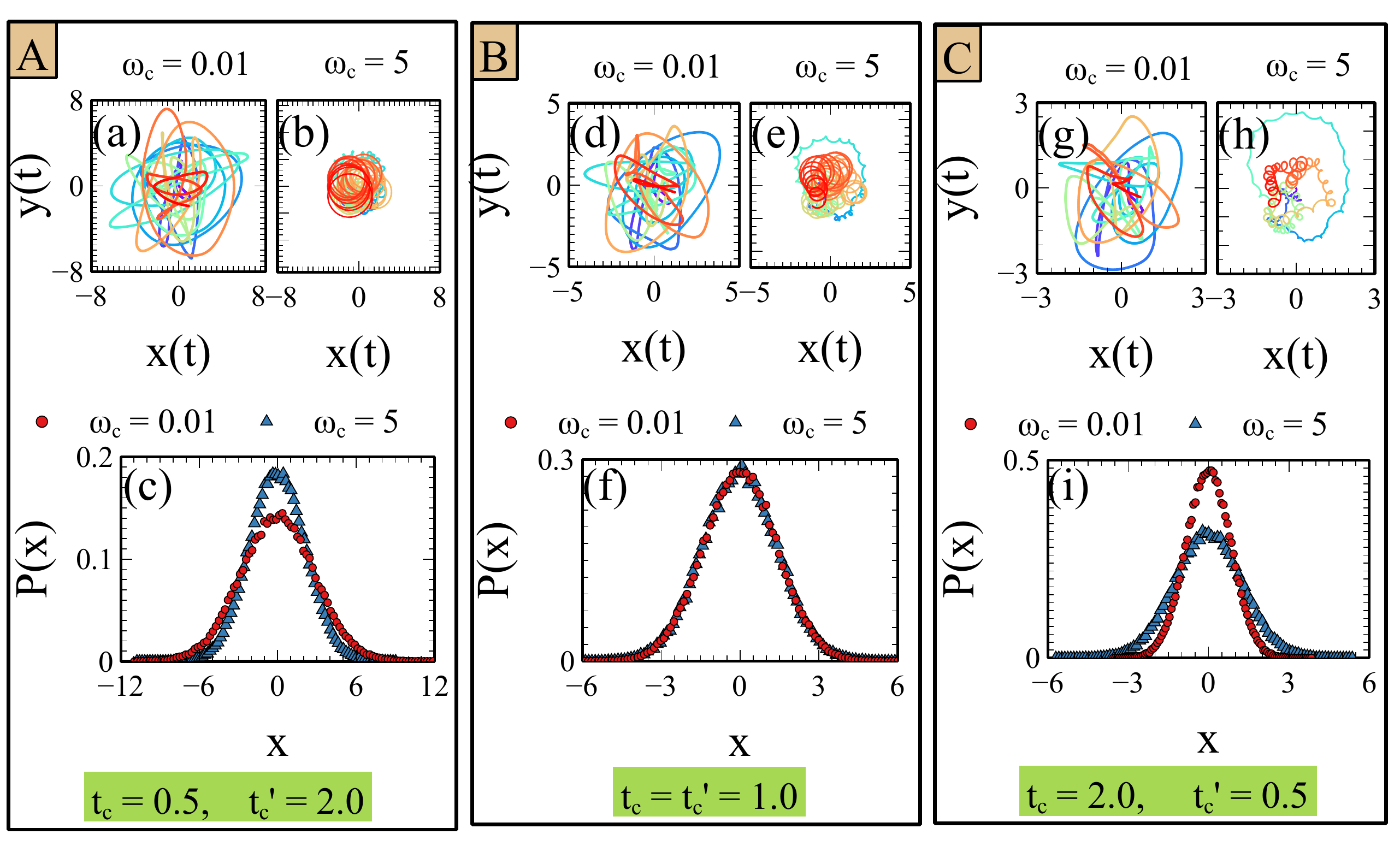}
    \caption{Particle trajectories for two different $\omega_{c}$ values and the corresponding steady state probability distribution functions are shown in (A), (B) and (C) for three different parameter regimes $t_{c} < t_{c}'$, $t_{c}=t_{c}'$ and $t_{c} > t_{c}'$, respectively. Particle trajectories in (a), (d) and (g) are for $\omega_{c}=0.01$ and in (b), (e) and (h) for $\omega_{c}=5.0$, respectively. For taking the trajectories, simulation is run for $t=100\;s$. The corresponding steady state position distributions are shown in (c), (f) and (i). The other common parameters are $\Gamma = 1.0$, $m = 1.0$, $\omega_0 = 1.0$ ,$v_0 = 1$, $z_0 = 0$, and $D = 1$.}
    \label{fig:traj_harm}
\end{figure*}
which represents a perfect spira mirabillis, familiar for active particles in Newtonian fluids \cite{sprenger2022active}. However, with increase in $t_{c}'$ circular motions are more stabilized and in  $t_{c}' \rightarrow \infty$ limit,  $\langle R(t)\rangle^{(f)}$ takes the form of a perfect circle of radius $\frac{|v_{0}|^{2}}{\omega_{c}^2}$ centred at $\frac{iv_{0}}{\omega_{c}}$. The same is reflected from the inset of Fig.~\ref{fig:md_free} (a), where $t_{c}^{'}$ is taken as $20$. With increase in $\omega_{c}$ value, the radius of the circular trajectories decreases and also centre of the circle shifts towards zero value, reflecting kind of trapping of the particle for high magnetic field [see Fig.~\ref{fig:md_free} (b)].

The 2D evolution of steady state MD with $\omega_{c}$ and $\Gamma$ are shown along with their corresponding parametric plots in Figs.~\ref{fig:md_free} (b) and (c), respectively. In the absence of magnetic field ($\omega_{c}=0$), the steady state MD is simply $\frac{2 v_{0}}{\Gamma}$. With increase in $\omega_{c}$, it decreases in the x-direction and increases in the y-direction and finally approaches to zero value for very large $\omega_{c}$ [Fig.~\ref{fig:md_free} (b)]. Similarly, for low value of $\Gamma$, where inertial influence is significant, the particle on average takes longer time to approach steady state and covers larger distances. However, the steady state is approached faster with increase in $\Gamma$ values as expected. The steady state MD starts from the value $\frac{iv_{0}}{\omega_{c}}$ for $\Gamma=0$ and approaches zero for very large $\Gamma$ value [Fig.~\ref{fig:md_free} (c)], where the inertial influence is negligibly small. This provides a delay in the dynamical behaviour of the particle because of inertia in the dynamics.

Using Eq.~\eqref{eq:msd-def}, we have exactly calculated the MSD in both transient and time asymptotic limit. In the time asymptotic limit ($t \rightarrow \infty$) or at steady state, the MSD is obtained as
\begin{equation}
   \langle R^2\rangle^{(f)}_{st} = \frac{8 D t}{m^2 \left(\Gamma ^2+4 \text{$\omega_c$}^2\right)}.
\end{equation}

From this expression, it is confirmed that $\langle R^2\rangle^{(f)}_{st}$ depends neither on $t_{c}^{'}$ nor on $t_{c}$. Hence, the overall distance covered by a free particle at steady state is completely independent of the underlying physical processes of the medium. However, it has dependence on $\omega_{c}$ and $\Gamma$. It varies inversely both with $\omega_{c}$ and $\Gamma$. From the transient behaviour of MSD, it is confirmed that motion is ballistic in the initial transient regimes as $\langle R^2\rangle \propto t^2$ and it is diffusive in long time regimes since $\langle R^2\rangle \propto t$. The intermediate time regimes of MSD represents kind of oscillatory behaviour, indicating the anomalous diffusion of the particle.

\subsection{Confined harmonic particle in the presence of magnetic field}\label{sec:harm-mag}

For a harmonically confined particle, the solution of the dynamics [Eq.~\eqref{eq:maineqmotion}] is given by
\begin{equation}
z(t) = \sum_{i=1}^{3}\left[ b_i z_0 e^{s_i t}+ a_i v_0 e^{s_i t} + a_i \int_{0}^{t} e^{s_i(t-t')}\xi(t') \,dt'\right].
\end{equation}
Here, $s_i$'s are the solution of the cubic equation
\begin{equation}
t_c^\prime s^3 + (1 - j\omega_ct_c^\prime)s^2  + \left( \frac{\Gamma}{2} - j\omega_c+ \omega_0^2t_c^\prime \right)s + \omega_0^2 = 0.
\end{equation}
$a_i$'s and $b_i$'s are given by the matrix equations
\begin{equation}
S\cdot A = \begin{pmatrix}
    0 \\ 
    -1 \\ 
    \frac{1}{t_c^\prime}
\end{pmatrix} \quad \text{and} \quad 
S\cdot B = \begin{pmatrix}
    1 \\ 
    -\frac{1}{t_c^\prime} + j\omega_c \\ 
    \frac{\Gamma}{2 t_c^\prime} -\frac{j\omega_c}{t_c^\prime}  
\end{pmatrix},
\end{equation}
with $A = \left[ a_i \right]_{3\times 1}$ and $B = \left[ b_i \right]_{3\times 1}$. The matrix $S$ is given by
\begin{equation}
S = \begin{pmatrix}
    1 & 1 & 1 \\
    s_2 + s_3  & s_1 + s_3 & s_1 + s_2 \\
    s_2s_3 & s_1s_3 & s_1s_2
\end{pmatrix}.
\end{equation}
\begin{figure}[!ht]
\includegraphics[scale=0.572]{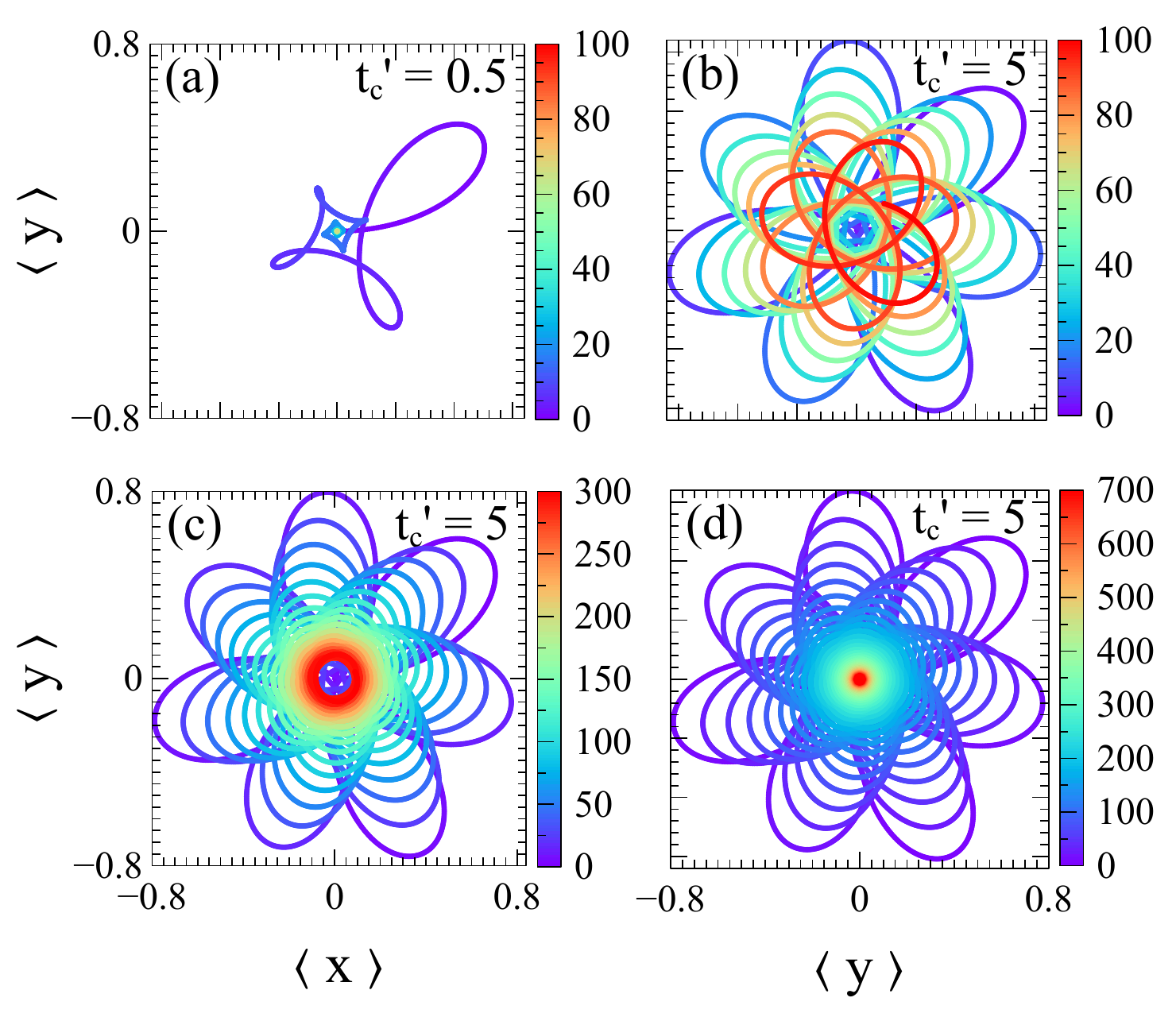}
\caption{For a harmonically confined particle ($\omega_{0}=1$), the 2D parametric plots of MD ($\langle y \rangle$ vs $\langle x \rangle$) [Eq.~\eqref{eq:md_harm}] are shown for $t_c' = 0.5$ in (a) and $t_c'=5.0$ in (b), (c) and (d), respectively. The color map shows the evolution of MD with time. The other common parameters are $\Gamma = 1.0$, $\omega_c = 1.0$, $m = 1.0$, $v_0 = 1$, $z_0 = 0$ and $D = 1$.}
\label{fig:md_harm}
\end{figure}

In order to understand the viscoelastic response of the particle to the magnetic field, we have shown the simulated particle trajectories of the dynamics [[Eq.~\eqref{eq:maineqmotion}]] and the corresponding steady state position distributions in Fig.~\ref{fig:traj_harm}. In harmonic confinement, the particle trajectories are circular and show interesting features with  magnetic field by the complex interplay between the corresponding time scales of elastic dissipation and active fluctuations in the medium. If the elastic dissipation persists for longer interval of time as compared to the duration of activity or self propulsion, i.e., for $t_{c}^{'} > t_{c}$ [see Fig.~\ref{fig:traj_harm} (a), (b) and (c)], the directed motion gets suppressed with increase in $\omega_{c}$ and the particle becomes incapable of taking advantage from the influence of magnetic field in making circular trajectories.  As a consequence, the circular trajectories get suppressed and the particle gets more confined towards the mean position of the well with increase in $\omega_{c}$ [Fig.~~\ref{fig:traj_harm} (b)]. The steady state position distribution becomes sharp across the centre or mean position with increase in $\omega_{c}$, deliberating the confinement of the particle with magnetic field [see Fig.~~\ref{fig:traj_harm} (c)]. If the activity persists for longer duration as compared to the elastic dissipation in the medium, i.e., for $t_{c} > t_{c}^{'}$ [see Fig.~\ref{fig:traj_harm} (g), (h) and (i)], the trajectories comparatively get enhanced with increase in $\omega_{c}$. The particle as if moves away from the centre initially and on average covers larger distances before coming back to the mean position of the well. That is why, the steady state position distribution gets broader and the peak of the distribution gets suppressed with increase in $\omega_{c}$, reflected from Fig.~\ref{fig:traj_harm} (i). 

However, if both elastic dissipation and activity persists almost for equal interval of time in the medium, i.e., for $t_{c}^{'} = t_{c}$, the steady state is not affected by the presence of magnetic field [see Fig.~\ref{fig:traj_harm} (d), (e) and (f)] and that is why the distribution shows same behaviour for both small and large $\omega_{c}$. In this limit, the generalized fluctuation dissipation relation (GFDR) is satisfied and the system approaches equilibrium. The MD [Eq.~\eqref{eq:md-def}] for a confined harmonic particle can be calculated as
\begin{equation}
    \langle R(t) \rangle^{(h)} = \langle z(t)-z(0) \rangle = \sum_{i=1}^{3}\left[ b_i z_0 e^{s_i t}+ a_i v_0 e^{s_i t}\right].
    \label{eq:md_harm}
\end{equation}
Expanding MD around $t = 0$, we get
\begin{equation}
    \langle R(t)\rangle^{(h)} = v_0 t+\frac{1}{2} \left(-\frac{\Gamma  z_0}{2 t_c'}+i v_0 \omega _c-z_0\omega _0^2 \right) t^2 +O\left(t^3\right).
\end{equation}

In the time asymptotic limit, $\langle R(t) \rangle^{(h)}$ approaches zero value implying that the particle comes back to the mean position of the well. In Fig.~\ref{fig:md_harm}, we have plotted the 2D parametric plot for the time evolution of MD for different values of $t_{c}^{'}$. For $t_{c}^{'}=0.5$, the particle almost take $100$ secs to approach steady state and come back to the mean position of the well as shown in Fig.~\ref{fig:md_harm}(a). However, with increase in $t_{c}^{'}$, the particle motion gets delayed and $100 secs$ is not enough for the particle to come back to the mean position as in Fig.~\ref{fig:md_harm}(b). With further increase in time, the particle approaches towards steady state [see Fig.~\ref{fig:md_harm}(c)] and finally for very long time, the particle comes back exactly to the mean position of the well [see Fig.~\ref{fig:md_harm}(d)]. An interesting feature of this behaviour is that the particle motion gets delayed due to the longer persistence of elastic dissipation in the medium, confirming a kind of delay in the dynamics due to presence of memory.

Using [Eq.~\eqref{eq:msd-def}], the MSD for harmonically confined particle can be calculated as
\begin{widetext}
\begin{equation}
\begin{split}
    \langle R^2(t) \rangle^{(h)} = & \left|\langle R(t) \rangle^{(h)}\right|^2  +\sum_{i=1}^3\sum_{j=1}^3 \frac{a_i a_j^* D}{m^2} \Biggl[ \frac{t_c e^{t \left(s^*_j-\frac{1}{t_c}\right)}}{\left(t_c s_i+1\right) \left(1-t_c s^*_j\right)}+\frac{t_c e^{t
   \left(s_i-\frac{1}{t_c}\right)}}{\left(1-t_c s_i\right) \left(t_c s^*_j+1\right)} \\
   & +\frac{t_c \left(s_i+s^*_j\right)-2}{\left(s_i+s^*_j\right) \left(t_c s_i-1\right) \left(t_c s^*_j-1\right)}+\frac{e^{t \left(s_i+s^*_j\right)}
   \left(t_c \left(s_i+s^*_j\right)+2\right)}{\left(s_i+s^*_j\right) \left(t_c s_i+1\right) \left(t_c s^*_j+1\right)} \Biggr].
\end{split}
\label{eq:msd_harm}
\end{equation}
\end{widetext}

\begin{figure}[!ht]
\includegraphics[scale=0.633]{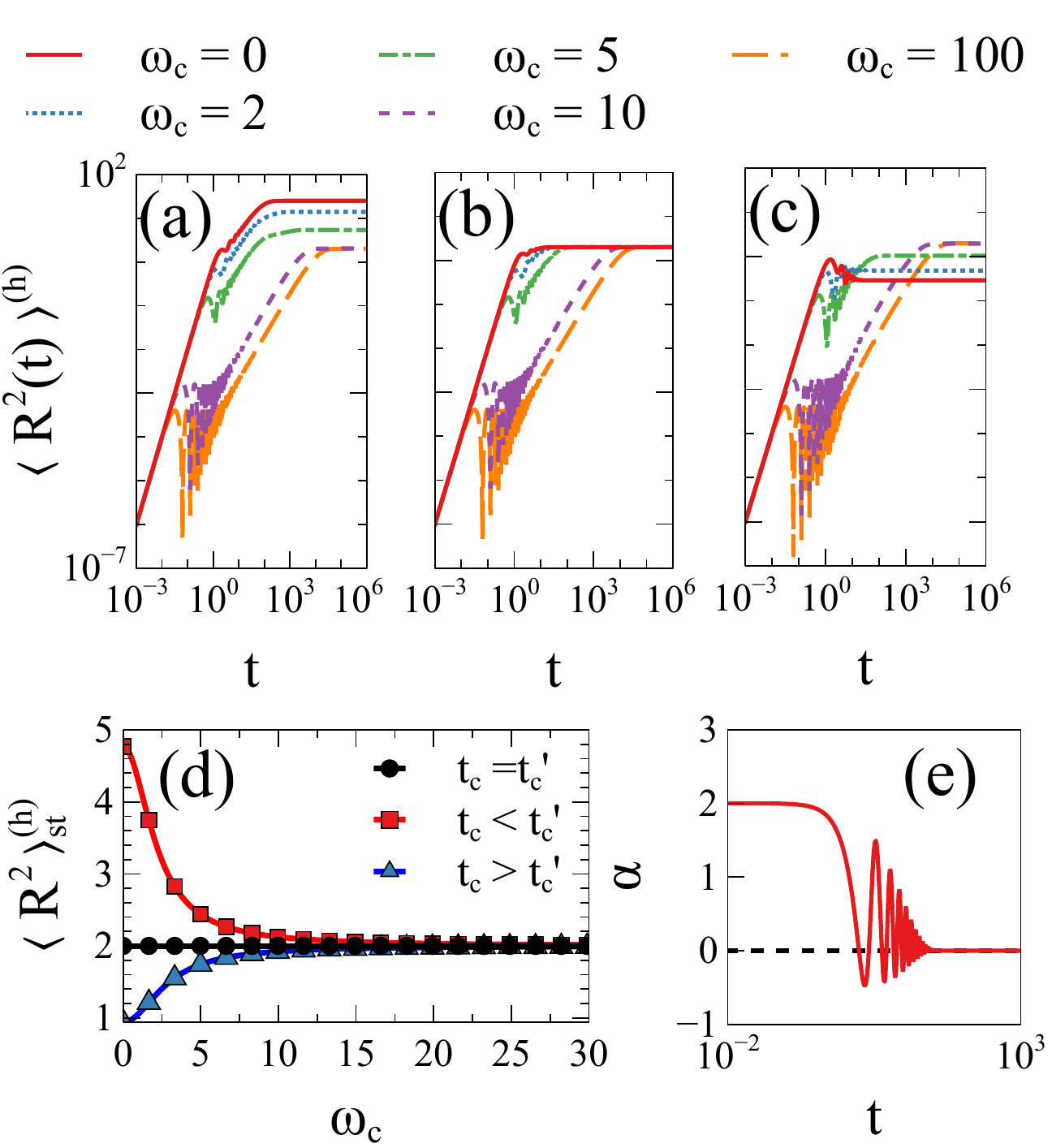}
\caption{For a harmonically confined particle ($\omega_{0}=1$), MSD as a function of $t$ [Eq.~\eqref{eq:msd_harm}] for different values of $\omega_{c}$ are shown for  $t_c < t_c'$ ($t_{c}=0.5$,$t_{c}'=2$) in (a), for $t_c = t_c' = 1$ in (b) and for $t_c > t_c'$ ($t_{c}=2$, $t_{c}'=0.5$) in (c), respectively. The steady state MSD [$\langle R \rangle^{(h)}_{st}$] [Eq.~\eqref{eq:msd_harm_steady}] is plotted as a function of $\omega_c$ in (d). The time variation of exponent ($\alpha$) is shown in (e) for $t_{c}=t_{c}'=1$. The other common parameters are  $m=1.0$, $\gamma =1.0$, $v_0 = 1$, $z_0 = 0$ and $D = 1$.}
\label{fig:msd_harm}
\end{figure}

In Fig.~\ref{fig:md_harm}, we have shown the plot of MSD as a function of $t$ for different values of $\omega_{c}$ in three different regimes, i.e., for $t_c < t_c'$, $t_c = t_c'$ and for $t_c > t_c'$. The MSD reflects the ballistic feature of the motion in the lower time regime as $ \langle R^2(t) \rangle^{(h)} \propto t^{2}$ and non-diffusive motion in the long time limit since $ \langle R^2(t) \rangle^{(h)}$ saturates to a constant value. In the intermediate regime of time, $\langle R^2(t) \rangle^{(h)}$ shows an oscillatory behaviour. The initial ballistic regimes always get suppressed with increase in $\omega_{c}$ irrespective of the value of $t_{c}$ and $t_{c}'$. However, the steady state MSD with $\omega_{c}$ shows rather complicated features due to the interplay between the physical processes of elastic dissipation and active fluctuations in the medium. It decreases with increase in $\omega_{c}$ when $t_c < t_c'$ [Fig.~\ref{fig:msd_harm}(a)], becomes independent of $\omega_{c}$ when $t_c = t_c'$ [Fig.~\ref{fig:msd_harm}(b)], and increases with $\omega_{c}$ when $t_c > t_c'$ [Fig.~\ref{fig:msd_harm}(c)]. Moreover, for a very large $\omega_{c}$ value, MSD approaches to the value in the equilibrium limit, where GFDR is satisfied, irrespective of both these time scales $t_{c}$ and $t_{c}'$[Fig.~\ref{fig:msd_harm}(d)] . In order to understand these interesting feature of steady state MSD, we have exactly calculated MSD in the stationary state. Because of the linearity of the model and Gaussian nature of noise, the steady state probability distribution is Gaussian and can be computed using the steady state correlation matrix method \cite{caprini2021inertial}. By introducing a correlation matrix $\Xi$, the steady state probability distribution can be expressed as
\begin{equation}
       P(w) = \frac{1}{(2\pi)^4 \sqrt{|\Xi|}}\exp\left[{-\frac{1}{2}\left( w^{T} \Xi^{-1} w \right)}\right],
    \label{eq:pdf_sol}
\end{equation}
with $w$ being a column vector. The exact form of $w$ is given in Appendix \ref{sec:corr_meth}. The elements of $\Xi$ are given by
\begin{equation}
    \left[\Xi_{i,j}\right] = \langle w_i w_j \rangle - \langle w_i \rangle \langle w_j \rangle.
    \label{eq:corr_mat}
\end{equation} 
We can find the solution for $\Xi$ using correlation matrix method formalism as described in Appendix \ref{sec:corr_meth}. From the elements of $\Xi$, the steady state MSD ($\langle R^2 \rangle^{(h)}_{st}$) can be obtained as
\begin{equation}
\langle R^2 \rangle^{(h)}_{st} = \Xi_{1,1} + \Xi_{2,2}.
\end{equation}
By simplifying the above equation, $\langle R^2\rangle^{(h)}_{st}$ can be expressed in the form
\begin{equation}
   \langle R^2\rangle^{(h)}_{st} = \langle R^2 \rangle_{E} + \langle R^2 \rangle_{NE},
   \label{eq:msd_harm_steady}
\end{equation}
where $\langle R^2 \rangle_{E}$  and $\langle R^2 \rangle_{NE}$ are the equilibrium and nonequilibrium contribution of MSD. $\langle R^2 \rangle_{E}$ is given by the expression
\begin{equation}
        \langle R^2\rangle_{E} = \frac{2D}{m^2\Gamma \omega_0^2}.
    \label{eq:msd_harm_steady_E}
\end{equation}
For $D=2\gamma k_{B} T$, $\langle R^2\rangle_{E}$ represents the MSD of a passive Brownian particle confined in a harmonic potential with $T$ as the temperature of the bath. The expression for $\langle R^2\rangle_{NE}$ is given by
\begin{widetext}
\begin{equation}
   \langle R^2\rangle_{NE} = \frac{4 D (t_c' - t_c) (t_c + t_c')^2 \left[ 2 t_c' + t_c(2 + t_c\Gamma + 2 t_c (t_c + t_c')\omega_0^2)\right]}{m^2\Gamma\left\{ \left[ 2 t_c' + t_c(2 + t_c\Gamma + 2t_c (t_c + t_c')\omega_0^2) \right]^2 + 4 t_c^2 (t_c + t_c')^2 \omega_c^2 \right\}}.
   \label{eq:msd_harm_steady_NE}
\end{equation}
\end{widetext}
From Eq.~\eqref{eq:msd_harm_steady_NE}, it is confirmed that $ \langle R^2\rangle_{NE}$ has a dependence on both $t_{c}$ and $t_{c}^{'}$. It is to be noted that $\langle R^2 \rangle_{NE}$ becomes zero when $t_{c}$ becomes equal to $t_{c}^{'}$. In this limit, we consider $D = 2\gamma k_BT$ for GFDR to be validated so that the system is in equilibrium. Also $\langle R^2 \rangle_{NE}$ vanishes in the presence of a very strong magnetic field since $\lim_{\omega_c \to \infty} \langle R^2 \rangle_{NE} = 0$. 

In both the above two cases, $\langle R^2 \rangle^{(h)}_{st}$ simply becomes $\langle R^2 \rangle_{E}$ which is independent of both $t_c$ and $t_c'$. However, Eq.~\eqref{eq:msd_harm_steady} indicate three interesting features of $\langle R^2 \rangle^{(h)}_{st}$ in the $t_c$-$t_c'$ parameter space, i.e., for $t_c < t_c'$, $t_c = t_c'$, and $t_c > t_c'$ regimes, respectively [Fig.~\ref{fig:msd_harm}]. For the parameter regime with $t_c < t_c'$, $\langle R^2 \rangle_{NE}$ in Eq.~\eqref{eq:msd_harm_steady_NE} is positive and hence $\langle R^2 \rangle^{(h)}_{st}$ is always greater than $\langle R^2 \rangle_{E}$. Thus, for the regime with $t_c < t_c'$, since $\langle R^2 \rangle_{NE}$ vanishes in $\omega_{c} \rightarrow \infty$ limit, the suppression of $\langle R^2 \rangle^{(h)}_{st}$ with $\omega_{c}$ is obvious. That is why, it decreases with increase in $\omega_c$ and approaches the equilibrium value in $\omega_{c} \rightarrow \infty$ limit. On the other hand, for the parameter regime with $t_c > t_c'$, $\langle R^2 \rangle_{NE}$ is negative and hence $\langle R^2 \rangle^{(h)}_{st}$ is less than its equilibrium value. Thus, enhancement of $\langle R^2 \rangle^{(h)}_{st}$ with $\omega_c$ is expected and that is what we notice in Fig.~\ref{fig:msd_harm}(d) that $\langle R^2 \rangle^{(h)}_{st}$ increases with increase in $\omega_c$ and saturates at the equilibrium value in the $\omega_{c} \rightarrow \infty$ limit. However, for the regime with $t_c = t_c'$, $\langle R^2 \rangle_{NE}$ vanishes and the steady state MSD is simply $\langle R^2 \rangle_{E}$. It approaches the equilibrium value for $D=2\gamma k_{B} T$ and becomes independent of $\omega_c$. In this limit, the $\langle R^2 \rangle^{(h)}_{st}$ is same as the case of a passive Brownian particle. Hence, for a finite magnetic field, the system displays a re-entrant type transition from active to passive and then to active behaviour with the elastic dissipation time scale of the dynamics. Further in order to understand the time dependence of $\langle R^2(t) \rangle^{(h)}$, we introduce a parameter $\alpha$ such that $\langle R^2(t) \rangle \sim t^\alpha$ and hence the time dependence of $\alpha$ can be expressed as

\begin{equation}
    \alpha = \frac{\partial \log \left(\langle R^2(t) \rangle\right)}{\partial \log t}.
    \label{eq:exponent}
\end{equation}
The variation of $\alpha$ with $t$ is shown in Fig.~\ref{fig:msd_harm}(e). $\alpha = 2$ confirms the ballistic motion for initial time regimes and $\alpha = 0$ implies the non-diffusive motion for long time regimes. The oscillatory behaviour of $\alpha$ with $t$ in the intermediate time regimes indicate the anomalous diffusion of the particle

\section{CONCLUSION}\label{sec:summary}
In this work, we have investigated the dynamics of an inertial charged active particle suspended in a viscoelastic medium with an exponentially decaying viscous kernel. The particle is further subjected to an external magnetic field of constant magnitude and in a direction perpendicular to the plane of its motion. The dynamics of the particle is modelled using the generalized Langevin equation with an exponentially correlated noise. We consider the elastic dissipation time scale and the noise correlation time scales are not of equal order. As a consequence, the GFDR is violated. By simulating the generalized Langevin equation and with the help of analytical calculations, we have exactly explored the transient as well as steady state response of both free and harmonically confined particle to an uniform magnetic field. One aspect of the transient dynamics is that the instantaneous mean displacement has dependence on the elastic dissipation time scale for both free and harmonic particles. The particle on average takes longer time to approach steady state because of the presence of memory in the viscous kernel. This is confirmed from the simulated particle trajectories as well as from the parametric plots or mean displacement calculations. 

However, the steady state response of a harmonically confined particle to a magnetic field shows interesting behaviour by complex interplay between the elastic dissipation time scale and activity or self propulsion time scale of the dynamics. In particular, if the elastic dissipation persists for longer interval of time as compared to the self propulsion or activity in the medium, the steady state MSD decreases with increase in magnetic field. On the other hand, if the self propulsion persists for longer time as compared to the elastic dissipation, the steady state MSD increases with increase in the magnetic field strength. In both the situations, for a very strong magnetic field the steady state MSD becomes independent of magnetic field and approaches the value at the equilibrium limit at which the GFDR is validated and the particle behaves like an inertial passive Brownian particle. Therefore, for a finite magnetic field, the system passes through a memory induced re-entrant type transition from active to passive and then to active behaviour. For a free particle, the steady state MSD always get reduced with increase in strength of magnetic field and approaches zero for a very strong magnetic field, reflecting field induced trapping of the particle. Finally, we believe that all of our aforementioned results are amenable for experimental verification. Our results are further important for controlling the self-propulsion or active dynamics of a particle by fine tuning the external magnetic field along with the physical properties of the suspension.

\section{Acknowledgement}
M.S. acknowledges the start-up grant from UGC Faculty recharge program, Govt. of India for financial support.

\appendix
\section{}\label{sec:corr_meth}
For a harmonically confined particle, the dynamics [Eq.~\eqref{eq:maineqmotion-vector}] can be expressed in the form of a vector equation
\begin{equation}
    \dot{w} = A\cdot w + B\cdot \eta,
    \label{eq:dynamics_matrix}
\end{equation}
with $w$ being the column vector
\begin{equation}
    w = \begin{pmatrix}
        x & y & v_x & v_y & u_x & u_y & \xi_x & \xi_y
    \end{pmatrix}^{T}.
\end{equation}
Here, $v_x$ and $v_y$ are the $x$ and $y$ components of velocity. The variables $u_x$ and $u_y$ are given by
\begin{equation}
    \begin{pmatrix}
        u_x \\
        u_y
    \end{pmatrix} = \int_{0}^{t} \Gamma\left(t-t'\right) {\bf \dot{X}}(t')\, dt'.
    \label{eq:ux_and_uy}
    \vspace{0.5em}
\end{equation}
The matrices $A$ and $B$ in Eq.~\eqref{eq:dynamics_matrix} are coefficient matrix and noise matrix, respectively. Using the correlation matrix method formalism, by introducing correlation matrix $\Xi$ as defined in Eq.~\eqref{eq:corr_mat}, the Eq.~\eqref{eq:dynamics_matrix} can take a new form
\begin{equation}
    A\cdot \Xi + \Xi \cdot A^T + B\cdot B^T = 0.
    \label{eq:Xi_relation}
\end{equation}
By solving Eq.~\eqref{eq:Xi_relation}, one can get the correlation matrix $\Xi$ with which the probability distribution [Eq.~\eqref{eq:pdf_sol}] can also be calculated. From the elements of the correlation matrix $\Xi$, the steady state MSD is calculated.

\end{document}